\documentclass[preprint]{aastex}
\usepackage{epsfig}
\newcommand{\Jy}{\mbox{Jy}}
\newcommand{\pc}{\mbox{pc}}
\shorttitle{Deep submillimetre survey of the Galactic Centre}
\shortauthors{Pierce-Price~et~al.}
\begin{document}
\title{A deep submillimetre survey of the Galactic Centre}

\author{D.~Pierce-Price\altaffilmark{1}, J.~S.~Richer\altaffilmark{1},
J.~S.~Greaves\altaffilmark{2}, W.~S.~Holland\altaffilmark{2},
T.~Jenness\altaffilmark{2}, A.~N.~Lasenby\altaffilmark{1},
G.~J.~White\altaffilmark{3,1}, H.~E.~Matthews\altaffilmark{4,2}, 
D.~Ward-Thompson\altaffilmark{5}, W.~R.~F.~Dent\altaffilmark{6},
R.~Zylka\altaffilmark{7}, P.~Mezger\altaffilmark{8},
T.~Hasegawa\altaffilmark{9}, T.~Oka\altaffilmark{10},
A.~Omont\altaffilmark{11} and G.~Gilmore\altaffilmark{12}}

\altaffiltext{1}{Cavendish Astrophysics, Cavendish Laboratory, Madingley Road,
  Cambridge, CB3 0HE, UK}
\altaffiltext{2}{Joint Astronomy Centre, 660 N. A`oh\=ok\=u Place,
  Hilo, HI 96720}
\altaffiltext{3}{Department of Physics, Queen Mary and
  Westfield College, University of London, Mile End Road, London E1 4NS,
  UK} 
\altaffiltext{4}{National Research Council of Canada, Herzberg
  Institute of Astrophysics, 5071 West Saanich Road, Victoria, BC V9E
  2E7, Canada}
\altaffiltext{5}{Department of Physics and Astronomy,
  Cardiff University, 5~The Parade, Cardiff CF2 3YB, Wales, UK}
\altaffiltext{6}{Astronomy Technology Centre,
  Blackford Hill, Edinburgh, EH9 3HJ, UK}
\altaffiltext{7}{ITA, Universit\"at Heidelberg, Tiergartenstrasse 15,
  D-69121 Heidelberg, Germany}
\altaffiltext{8}{Max-Planck-Institut f\"ur
  Radioastronomie, Auf dem H\"ugel 69, D-53121 Bonn, Germany}
\altaffiltext{9}{Institute of Astronomy, University of
  Tokyo, 2-21-1 Osawa, Mitaka, Tokyo 181-0015, Japan}
\altaffiltext{10}{Department of Physics, Graduate School of Science,
  The University of Tokyo, 7-3-1 Hongo, Bunkyo-ku, Tokyo 113-0033, Japan}
\altaffiltext{11}{Institut d'Astrophysique de Paris, CNRS, 98bis Boulevard
  Arago, F 75014 Paris, France}
\altaffiltext{12}{Institute of Astronomy,University of Cambridge,
  Madingley Road, Cambridge. CB3 0HA, UK}

\begin{abstract}
  
  We present first results from a submillimetre continuum survey of
  the Galactic Centre `Central Molecular Zone' (CMZ), made with SCUBA
  on the James Clerk Maxwell Telescope. SCUBA's scan-map mode has
  allowed us to make extremely wide-field maps of thermal dust
  emission with unprecedented speed and sensitivity. We also discuss
  some issues related to the elimination of artefacts in scan-map
  data. Our simultaneous 850/450$\,\micron$ maps have a total size of
  approximately $2.8\times0.5\arcdeg$ ($400\times75\,\pc$) elongated
  along the galactic plane. It covers the Sgr~A region---including
  Sgr~A*, the circumnuclear disc, and the $+20\,{\rm km}/{\rm s}$ and
  $+50\,{\rm km}/{\rm s}$ clouds;
  the area around the Pistol; Sgr~B2---the brightest feature on the
  map; and at its Galactic Western and Eastern edges the Sgr~C and
  Sgr~D regions. There are many striking features such as filaments
  and shell-like structures, as well as point sources such as Sgr~A*
  itself. The total mass in the Central Molecular Zone is greater than
  that revealed in previous optically-thin molecular line maps by a
  factor of $\sim 3$, and new details are revealed on scales down to
  $0.33\,\pc$ across this $400\,\pc$-wide region.

\end{abstract}


\section{Introduction}

The Galactic Centre provides our best opportunity to study the
astrophysical processes in a galactic nucleus with high spatial
resolution. These include star-formation and the molecular cloud
structures in a region subjected to strong shear forces, magnetic
fields, and gravitational potentials. Thermal dust continuum emission
traces the temperature-weighted column density of material in a less
biased way than molecular line maps, which are affected by excitation,
optical depth, and abundance variations. Thus, we can use the
continuum emission to map the mass distribution in molecular clouds
with good accuracy and sensitivity.

We used SCUBA \markcite{1999MNRAS.303..659H}({Holland} {et~al.} 1999)
on the 15-m James Clerk Maxwell Telescope (JCMT), between April 1998
and April 2000, to map the Galactic Centre region in the submillimetre
continuum at $450\,\micron$ and $850\,\micron$. The new images cover a
region 4 times larger and $\sim 100$ times deeper than earlier surveys
\markcite{1994ApJ...424..189L}({Lis} \& {Carlstrom} 1994). The map
size is approximately $2.8\times0.5^\circ$, or $400\times75\,\pc$ at
an assumed distance of $8.5\,{\rm kpc}$. The map covers the entire
`Central Molecular Zone' (CMZ), which extends to radii of $200\,\pc$
and contains up to $10\%$ of the Galaxy's molecular interstellar
medium (ISM) \markcite{1996ARA&A..34..645M}({Morris} \& {Serabyn}
1996).

\section{Observations and data reduction}\label{sec:observ-data-reduct}

SCUBA \markcite{1999MNRAS.303..659H}({Holland} {et~al.} 1999) is a
submillimetre continuum bolometer camera with two arrays, one of 91
pixels optimised for $450\,\micron$, and one of 37 pixels optimised
for $850\,\micron$. In use, both arrays observe simultaneously. The
resolution is $8\arcsec$ FWHM at $450\,\micron$ and $15\arcsec$ at
$850\,\micron$.

SCUBA was used in scan-map mode, in which the array is scanned across
the sky at a rate of $24\arcsec$ per second whilst chopping the
secondary mirror at $7.8\,{\rm Hz}$. This chopping produces images in
which the sky is convolved with the primary beam of the telescope and
a dual-beam function, of positive and negative delta functions
separated by the chop throw.

The majority of the map was observed six separate times, with chop
throws of 20, 30, and $65\arcsec$ in both R.A. and Dec., although some
fields at the Galactic Eastern and Western ends were observed with 30,
44, and $68\arcsec$ chops. Each dual-beam image contains no
information at spatial frequencies equal to the inverse chop throw and
its harmonics, and the constant power level (zero spatial frequency)
is also lost.

The data were reduced with SURF, the standard SCUBA reduction software
\markcite{SUN216}({Jenness} \& {Lightfoot} 1999), using the ``Emerson
2'' deconvolution algorithm. This algorithm combines the multiple
chop-throw images of the map fields. A weighted average of these
observations is formed in Fourier space, and the reverse transform
yields the final map \markcite{1998adass...7..216J}({Jenness} \&
{Lightfoot} 1998). Although the combination of multiple chops improves
the spatial frequency coverage, it is unavoidable that spatial scales
more extended than a few times the maximum chop-throw are measured
with reduced sensitivity, and the zero spatial frequency is still
lost. Ultimately, the survey is sensitive to spatial scales in the
approximate range $0.3\,\pc$ (the beam size at $450\,\micron$) to
$\sim 10\,\pc$ (a few times the maximum chop-throw).

Calibration was performed from maps of Uranus and Mars, and all data
were corrected for atmospheric extinction using skydips. Zenith
opacities were in the range 0.1--0.45 at $850\,\micron$ and 0.33--2.5
at $450\,\micron$. Most of the data in the central $1.5\arcdeg$ region
was observed in extremely good weather, with $\tau(850\,\micron) <
0.13$, $\tau(450\,\micron) < 0.5$.

To calibrate fluxes of extended structures, the integrated signal in a
$1\arcmin$ diameter aperture was used to calculate a conversion factor
in $\Jy\,{\rm arcsec}^{-2}\,$V$^{-1}$. Calibration relative to the
primary calibrators is estimated to be accurate to 5\% at
$850\,\micron$ and 20\% at $450\,\micron$, based on the dispersion of
measured flux conversion factors, although absolute cloud fluxes also
depend on corrections for constant map offsets as described in section
\ref{sec:corr-artef}.

The survey has resolutions of approximately $8\arcsec$ at
$450\,\micron$ and $15\arcsec$ at $850\,\micron$, corresponding to
distances of $0.33\,\pc$ and $0.62\,\pc$ at a distance of $8.5\,{\rm
  kpc}$. The $1$-$\sigma$ sensitivities per beam are approximately
$30\,{\rm mJy}$ and $300\,{\rm mJy}$ at $850\,\mu{\rm m}$ and
$450\,\mu{\rm m}$ respectively, or $\sim20\,M_\sun$ per beam with the
assumptions in section \ref{sec:cloud-masses}.

\subsection{Corrections for artefacts}\label{sec:corr-artef}

Making large-scale SCUBA scan-maps is difficult for several reasons.
The total power level is not recorded, so there may be a constant
offset on each field, and this offset may in general vary between
fields in the mosaic. Within individual fields, one must account for
baseline removal on the time-series signals from the bolometers. The
dual-beam chop will suppress information at certain spatial
frequencies, as described above, and as there is so much extended
structure in the CMZ it is difficult to scan or chop onto regions of
no emission.

Some previous SCUBA scan-maps have sometimes exhibited reduction
artefacts, such as negative troughs around bright sources. We have
found that negative troughs can be greatly reduced by careful baseline
removal on a per-bolometer basis. A single $10\arcmin\times10\arcmin$
map is contructed from typically 10--15 scans of the array across the
field. Calculating a baseline separately for each scan will produce
large spurious offsets for those scans which pass over bright
structure, so ideally it is best to determine a baseline using only
those scans which contain little bright emission. This proved
impossible for a region as rich in structure as the Galactic Centre,
but we have found that using the median level over all the scans for
each bolometer does produce good results without the spurious offsets.

This is the largest area ever scan-mapped at 450 and $850\,\micron$,
with more than fifty overlapping $10\arcmin\times10\arcmin$ fields,
which must be mosaiced to form the final image. The Starlink CCDPACK
package \markcite{SUN139}({Draper}, {Taylor}, \& {Allan} 2000) was
used to combine the dual-beam images and remove relative level offsets
between fields of order 10 and $0.5\,{\rm mJy}\,{\rm arcsec}^{-2}$ at
450 and $850\,\micron$ respectively. We then applied further
corrections to ensure that the pixel mean calculated along the chop
throw direction was zero, for all of the mosaiced dual-beam maps. This
must be true, as long as the edges of the map are essentially
emission-free, and is necessary because incorrect zero levels in the
mosaiced dual-beam maps will translate to slopes in the final images.
We finally corrected for the constant zero offset on the entire map by
measuring the typical zero levels from map areas devoid of
identifiable clouds, and assigning an uncertainty based on the level
variations. The final uncertainty in baseline levels is $15$ and
$1\,{\rm mJy}\,{\rm arcsec}^{-2}$ at 450 and $850\,\micron$
respectively, the former being higher in part due to the greater
calibration uncertainties. These uncertainties have been neglected,
compared to the sensitivity noise levels of the maps.

\section{Analysis}

\subsection{General features}\label{sec:general-features}

\begin{figure}
\centerline{\epsfig{file=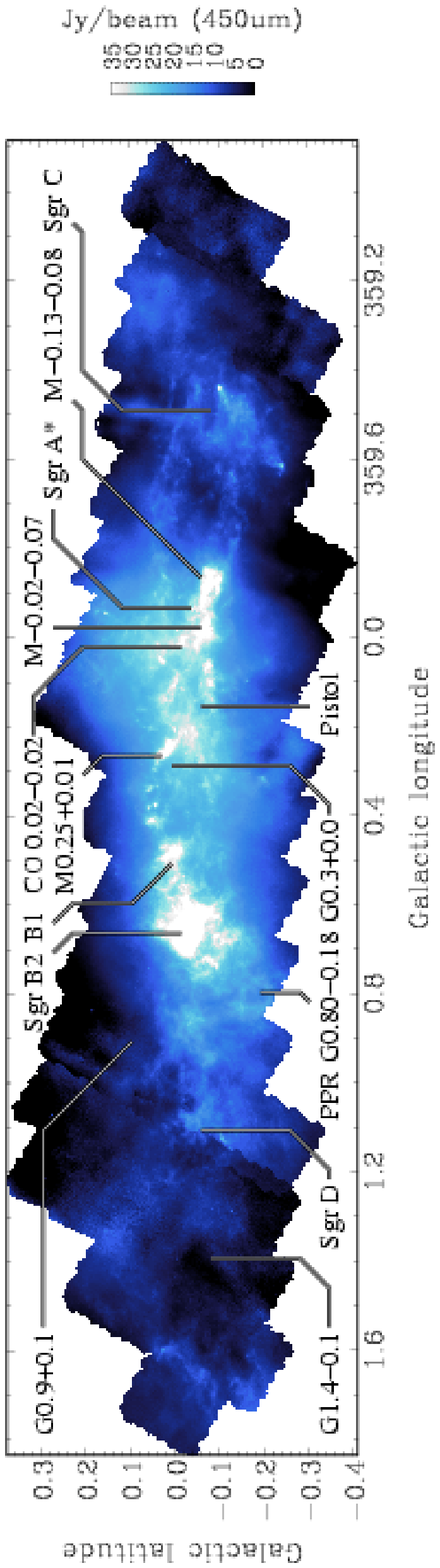,height=20cm,clip=}} \caption{SCUBA Galactic Centre Survey: $450\,\micron$ flux density.\label{fig:sho}}
\end{figure}

\begin{figure}
  \centerline{\epsfig{file=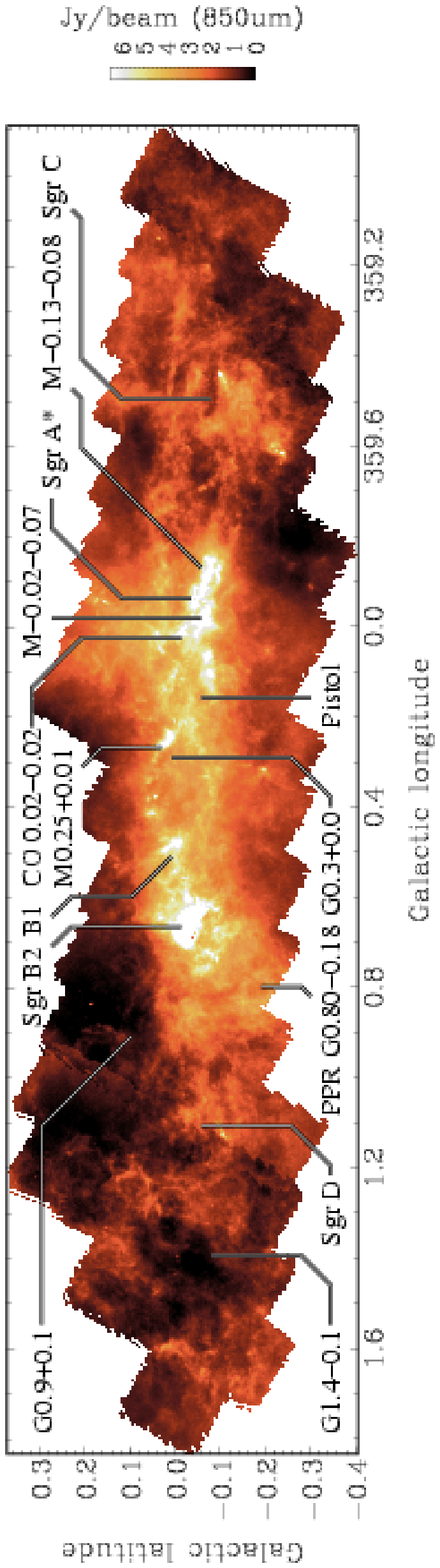,height=20cm,clip=}} \caption{SCUBA Galactic Centre Survey:
    $850\,\micron$ flux density.\label{fig:lon}}
\end{figure}

Our maps (figures \ref{fig:sho} and \ref{fig:lon}) show good
correspondence with previous surveys of the region, such as that by
\markcite{1994ApJ...424..189L}{Lis} \& {Carlstrom} (1994) but with
much higher sensitivity and resolution. At $850\,\micron$ we can
clearly see the central source Sgr~A* and its Circum-Nuclear Disc
(CND) ($l=-0.05$, $b=-0.05)$), although these features are less
pronounced at $450\,\micron$. We also see the compact GMCs M-0.13-0.08
(the $+20\,{\rm km}/{\rm s}$ cloud) and M-0.02-0.07 (the $+50\,{\rm
  km}/{\rm s}$ cloud).  Close to this is the high-velocity molecular
cloud CO 0.02--0.02 \markcite{1999ApJ...515..249O}({Oka} {et~al.}
1999). At higher galactic longitude we see the Pistol region
($l=0.15$, $b=-0.05$), other GMCs such as M0.25+0.01
\markcite{1994ApJ...423L..39L}({Lis} {et~al.} 1994), Sgr~B1 ($l=0.5$)
and B2 ($l=0.66$), and the Sgr~D ($l=1.1$) star-forming region. At
more negative longitudes, the map extends to the Sgr~C ($l=-0.5$) star
forming region.  All of the field is rich in extended structure,
dominated by filamentary clouds and cavities which may be associated
with supernova remnants.  Such structures have also been observed by
\markcite{1998ApJS..118..455O}{Oka} {et~al.} (1998) in CO ($J =
1$--$0$) emission.

\subsection{Sagittarius A*}

At both wavelengths we have a clear detection of the Sgr~A*
point-source.  By taking slices through the position of Sgr~A* to
determine and remove the local background, we estimate the flux from
Sgr~A* itself, with quoted 1-$\sigma$ uncertainties. The results are
also subject to the overall calibration uncertainties described in
section \ref{sec:observ-data-reduct}.  At $850\,\micron$ we measure a
flux of $2.6 \pm 0.3\,\Jy$. This is consistent with the value measured
from CSO-JCMT interferometry \markcite{1997ApJ...490L..77S}({Serabyn}
{et~al.} 1997) of $3.2 \pm 0.7\,\Jy$. At $450\,\micron$, the same
procedure gives a flux for Sgr~A* of $1.2\pm0.4\,\Jy$.  This is
consistent with the upper limit of $1.5\,\Jy$ found by
\markcite{1993ApJ...410..650D}{Dent} {et~al.} (1993). These fluxes are
somewhat lower than those measured by
\markcite{1995A&A...297...83Z}{Zylka} {et~al.} (1995), which were
$3.5\pm0.5\,\Jy$ at $800\,\micron$ (rather than $850\,\micron$) and
$3.0\pm1.0\,\Jy$ at $450\,\micron$, suggesting that there may be some
variability of the Sgr~A* point source. For example,
\markcite{1999cpg..conf..105T}{Tsuboi}, {Miyazaki}, \& {Tsutsumi}
(1999) observed a flare at 3mm during which the flux doubled, and
\markcite{1997ApJ...490L..77S}{Serabyn} {et~al.} (1997) found a flux
of $7 \pm 2\,\Jy$ at $350\,\micron$. Our fluxes indicate a spectral
index for the emission from Sgr~A* of $-1.5_{-1.4}^{+1.0}$, consistent
with a non-dust component in the emission. Flux values for the
synchrotron component from the postulated black hole are discussed
further by \markcite{2000ApJ...534L.173A}{Aitken} {et~al.} (2000).

\subsection{Greybody fitting}\label{sec:greybody-fitting}

Although we can calculate spectral indices in the submillimetre regime
from the SCUBA data, these data do not constrain the overall form of
the dust emission, and do not allow us to estimate the temperature of
the dust. To do this, we need data around the peak of the emission,
which for dust at these temperatures is in the far infra-red.

We have combined our submillimetre data with $100\,\micron$ data from
the IRAS Galaxy Atlas \markcite{1997ApJS..111..387C}({Cao} {et~al.}
1997), and fitted a single temperature greybody spectrum of the form
$I_\nu = (1-\exp(-\tau))B_\nu(T)$, where the optical depth $\tau$ is
expressed in terms of an index $\beta$ and a critical frequency
$\nu_{\rm crit}$ at which the emission becomes optically thick: $\tau
= ({\nu}/{\nu_{\rm crit}})^{\beta}$. Although the IRAS Galaxy Atlas
also contains $60\,\micron$ data, we believe that emission at this
wavelength traces a higher temperature component. By fitting to the
$850\,\micron$, $450\,\micron$, and $100\,\micron$ points we calculate
properties of the cold dust only.

We have performed a least-squares greybody fit to the three datasets
over the survey area, fitting for temperature $T$ and critical
frequency $\nu_{\rm crit}$. We fixed $\beta=2$, the theoretical
maximum for crystalline dust grains. Since the resolution of the IRAS
images was approximately $2\arcmin$, the SCUBA data were also smoothed
to this resolution.

The fits indicate a fairly uniform dust temperature of $\sim
21\pm2\,{\rm K}$. There are no signs of strong temperature variations
correlated with individual clouds, and this is further evidence that
the brightness structure in the maps is dominated by column density
variations. Therefore, we adopt a typical temperature of $20\,{\rm K}$
throughout the rest of this paper.

If $\beta$ is allowed to vary, the fit is better but the derived
temperature does not significantly change. Best-fit values of $\beta$
are $\beta \sim 2.4$, higher than the canonical value of $\beta=2$.
There is now much evidence for $\beta > 2$:
\markcite{1998ApJ...507..794L}{Lis} \& {Menten} (1998) and
\markcite{1999cpg..conf..453D}{Dowell} {et~al.} (1999) measured $\beta
\sim 2.8$ and $\beta \sim 2.5$ respectively in GCM0.25+0.01.  In other
regions, \markcite{1998ApJ...509..299L}{Lis} {et~al.} (1998) measure
$\beta$ values as high as $2.5$ in the Orion Ridge and
\markcite{1999A&A...347..640B}{Bernard} {et~al.} (1999) also report
PRONAOS observations of high $\beta$ in the Polaris cirrus cloud. In
laboratory work \markcite{1995ApJ...446..902K}{Koike} {et~al.} (1995)
report $\beta$ up to $2.7$ in carbon-based compounds at room
temperature, and \markcite{1996ApJ...462.1026A}{Agladze} {et~al.}
(1996) have shown $\beta > 2$ in silicates between $\approx 5$ and
$20\,{\rm K}$.

\subsection{Spectral index map}

\begin{figure}
  \centerline{\epsfig{file=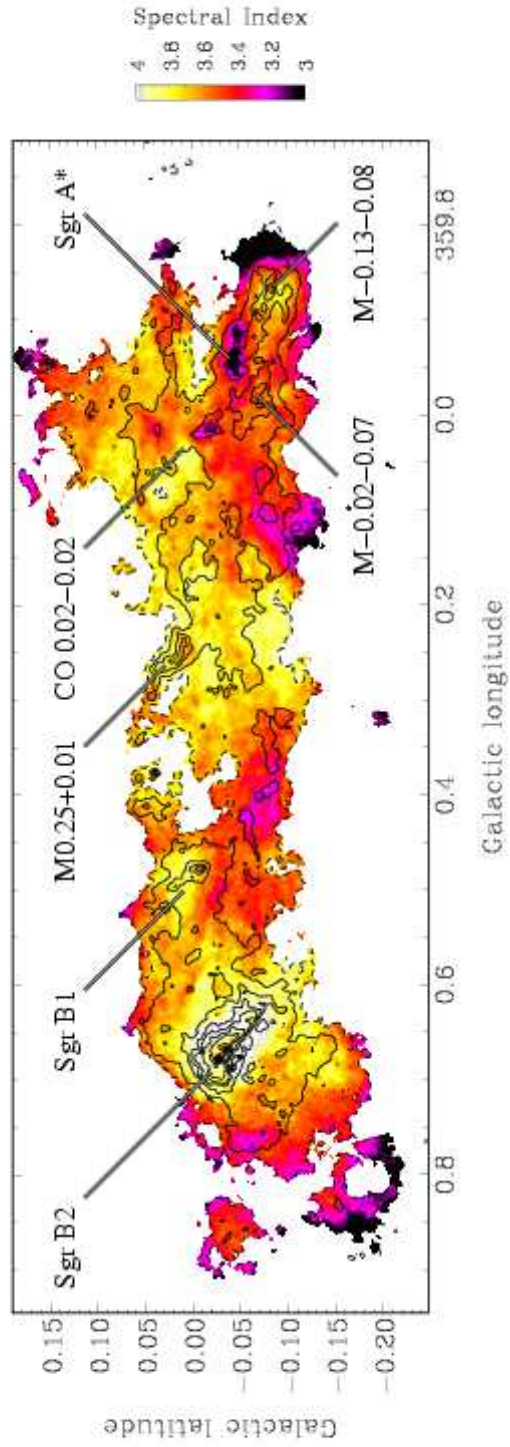,height=20cm,clip=}} \caption{Spectral index map for central region,
    with contours of $850\,\micron$ map.\label{fig:alpha}}
\end{figure}

We have used our $450\,\micron$ and $850\,\micron$ maps to generate a
map of spectral index $\alpha$ (defined by $S_{\nu} \propto
\nu^{\alpha}$), which is shown in figure \ref{fig:alpha}.  We first
smoothed the $450\,\micron$ map to the lower resolution of the
$850\,\micron$ observations, and imposed a lower threshold on both
maps of $\sim 3\sigma$, since spectral indices derived from small flux
values are subject to large errors.

Over most of the bright parts of the map $\alpha$ is in the range
3--4, indicating rather constant dust properties with a dust opacity
index $\beta \sim 2$. The bulk spectral index for M-0.02-0.07 (the
$+50\,{\rm km}/{\rm s}$ cloud) is $\alpha = 3.5 \pm 0.05$, whereas for
M-0.13-0.08 (the $+20\,{\rm km}/{\rm s}$ cloud) it is $\alpha = 3.7
\pm 0.05$ (quoted errors are from the variation within the clouds).
Assuming greybody emission at a dust temperature of $20\,{\rm K}$,
these indicate dust opacity indices of $\beta \sim 2.2$--$2.4$.
Although $\beta=2$ is the theoretical upper limit for crystalline dust
grains, in section \ref{sec:greybody-fitting} we discuss some evidence
for higher values of $\beta$.

Differences in spectral index may be due to differences in
temperature. Since at these temperatures, emission at $450\,\micron$
and $850\,\micron$ is not truly in the Rayleigh-Jeans part of the
spectrum, the difference between the spectral index $\alpha$ and the
dust opacity index $\beta$ is temperature dependent. It is therefore
possible that two clouds may have the same $\beta$ but different
temperatures, leading to different spectral indices. If this were the
case, a lower $\alpha$ would indicate a cooler cloud. This cannot
account for all the $\alpha$ variations: for example, $\alpha$ is {\em
  low} at only $3.2\pm0.1$ for CO 0.02-0.02, and this cloud is thought
to be relatively {\em hot} at $\sim 60\,{\rm K}$
\markcite{1999ApJ...515..249O}({Oka} {et~al.} 1999).  In section
\ref{sec:greybody-fitting}, we detect no significant temperature
variations on $\sim 2\arcmin$ scales.

\subsection{Cloud masses}\label{sec:cloud-masses}

To estimate the total cloud masses from their integrated fluxes, we
require the grain absorption cross-section per unit mass of gas and
dust. We have adopted the parameters suggested by
\markcite{1994ApJ...421..615P}{Pollack} {et~al.} (1994) with opacity
index $\beta = 2$, and a correction for Galactic Centre metallicity
$Z/Z_\sun = 2$ \markcite{1989A&A...209..337M}({Mezger} {et~al.} 1989).
We thus derive opacities of $\kappa_{450\mu{\rm m}} =
4.94\times10^{-3}\,\rm{m}^2\,\rm{kg}^{-1}$ and $\kappa_{850\mu{\rm m}}
= 1.38\times10^{-3}\,\rm{m}^2\,\rm{kg}^{-1}$.  Assuming a uniform dust
temperature of $20\,{\rm K}$ we have for $\lambda = 450$ and
$850\,\micron$, conversion factors of 63.3 and 513 $M_\sun / \Jy$
respectively.  The total mass found by the survey, measured by
integrating the total $850\,\micron$ brightness, is then
$(53\pm10)\times10^6 M_\sun$.

The largest source of error in these mass measurements is our
uncertainty of the dust temperature. The value of $20\,{\rm K}$ is
derived in section \ref{sec:greybody-fitting}, but
\markcite{1991ApJ...380..429L}{Lis}, {Carlstrom}, \& {Keene} (1991)
adopt a mean temperature of $30\,{\rm K}$.  However, the ISO study of
GCM0.25+0.01 by \markcite{1998ApJ...507..794L}{Lis} \& {Menten} (1998)
gives a mean temperature of $\sim 18\,{\rm K}$ for this core,
suggesting that the lower dust temperatures are correct.  We neglect
optical depth effects as the emission is optically thin at
$850\,\micron$ except from Sgr~B2.

\section{Further discussion}\label{sec:further-discussion}

The improved depth, resolution, and extent of the SCUBA survey reveal
many new features in the Galactic Centre. Not only are there the known
large star-forming cloud complexes but also a wide-ranging network of
dusty filaments is revealed. For example, at negative Galactic
longitudes close to $b=0$ there is filamentary structure which may be
continuous and stretches along the Galactic plane for up to
$0.8\arcdeg$ ($120\,\pc$). There are partial shells, possibly
associated with the known SNRs Sgr~A East (G0.0+0.0), G0.3+0.0,
G0.9+0.1, and G1.4-0.1 \markcite{1998cgsr.book.....G}({Green} 1998).
There is also a circular shell of dust clouds centred on $l,b =
(0.8,-0.18)$ which does not appear to be documented, and which we have
labelled PPR~G0.80-0.18.  This may trace a wind-blown region or
perhaps a supernova remnant. In a future paper, we will contrast these
structures with star-forming regions near the Solar circle, comparing
masses, densities, dust grain properties and star-forming efficiency.

The total mass detected by our survey is $(53\pm10)\times10^6 M_\sun$,
in the inner $400\,\pc$. \markcite{1998A&A...331..959D}{Dahmen}
{et~al.} (1998) derived a weighted best estimate of the total mass in
the inner $600\,\pc$ of the Galactic centre using a combination of
dust emission, gamma rays, and C$^{18}$O molecular line emission. This
estimate was $20$--$50\times10^6 M_\sun$. The best optically-thin
tracer in this set is the C$^{18}$O survey, from which was derived a
mass of $17\times10^6 M_\sun$ in the central $600\,\pc$. Although our
survey covers a smaller area, the total mass detected is significantly
greater. Therefore, the SCUBA data comprise the first optically-thin
map to trace essentially all the mass in the CMZ at high resolution.

\acknowledgments We are grateful to the UK Panel for Allocation of
Telescope Time for the award of observing time for this project. JSR
acknowledges a Royal Society Fellowship and DP-P a PPARC Research
Studentship. The JCMT is operated by the Joint Astronomy Centre on
behalf of PPARC of the UK, the Netherlands OSR, and NRC Canada. We
acknowledge the support provided by the Starlink Project which is run
by CCLRC on behalf of PPARC. We also would like to thank an anonymous
referee for some very useful comments.




\end{document}